\documentclass[letterpaper,american,reprint, aps, prl, twocolumn]{revtex4-2}
\usepackage[T1]{fontenc}
\usepackage[latin9]{inputenc}
\usepackage{units}
\usepackage{amstext}

\makeatletter

\pdfpageheight\paperheight
\pdfpagewidth\paperwidth

\makeatother

\usepackage{babel}
\begin{document}
\title{Observing evolution from steady state}
\author{Herman Telkamp}
\address{Jan van Beverwijckstraat 104, 5017JA Tilburg, The Netherlands}
\email{herman_telkamp@hotmail.com}

\begin{abstract}
\noindent The time-translation symmetry of the conformal FLRW frame
$\bar{g}=a^{-2}g$ allows reinterpretation of cosmological observation
in the static space of a stationary universe, where constant matter
density $\bar{\rho}_{\textrm{m}}=\rho_{\textrm{m0}}$ induces constant
curvature $R_{0}^{-2}$. A hyperbolic de Sitter solution arises from
equipartition of the kinetic energy of recessional and peculiar components
of the gravitational field, corresponding to a total density $24R_{0}^{-2}$
of twice the scalar curvature. This predicts a matter density $\Omega_{\textrm{m}}=\nicefrac{1}{24}$,
or a Hubble constant $h=\sqrt{24\rho_{\textrm{m0}}}\approx$$0.72$,
in agreement with distance-ladder estimates. Projecting the equilibrium
state onto the $\Lambda\textrm{CDM}$ model returns $\hat{h}=\frac{4}{3}\sqrt{12\rho_{\textrm{m0}}}\approx0.68$
and exact densities $\hat{\Omega}_{\textrm{m}}=1-\hat{\Omega}_{\Lambda}=[\textrm{sinh}(\frac{4}{3}\textrm{asinh}(1))^{2}+1]^{-1}=0.3179...$,
within confidence limits of Planck 2018 results.
\end{abstract}
\maketitle
Invariance of de Sitter universe with respect to expansion of the
scale factor is physically attributed to some unknown constant energy
density. In an essay of 1906 on the relativity of space, Poincar\'e
however imagined invariance of an expanding universe as arising from
uniform spatial expansion, that is, expansion without exception of
gravitationally bound objects, so that everything expands, including
photon wavelength and ruler, but nothing seems to change \citep{PoincareScienceMethod}.
The uniformly expanding universe thus shows the time-translation symmetry
of de Sitter universe, while the constant matter density can be related
to constant curvature $R_{0}^{-2}$. 

The expansion of physical units of length makes space in the frame
of the comoving observer static. In this maximally symmetric frame
one expects the universe to be infinitely old and in thermodynamic
equilibrium, meeting the perfect cosmological principle. Repeated
observation of cosmological parameters, like density or temperature,
will show no change, hence zero redshift drift, as evidence of a stationary
state in a static space. Still, in terms of the \textsl{present} unit
of length, the universe continues to expand from past into future,
as revealed by cosmological redshift on the past light cone. The same
universe then appears evolving and of finite age, like in Big Bang
cosmology. Poincar\'e's uniformly expanding universe thus seems to
unify aspects of Big Bang and steady state scenarios \citep{Hoyle,BondiGold}. 

Uniform expansion can be represented conformally by the metric $\bar{g}=a^{-2}g$,
where $a(t)$ is the scale factor and $g$ is the FLRW metric. In
standard coordinates, the line element of $\bar{g}$ is
\begin{equation}
d\bar{s}^{2}=a^{-2}dt^{2}-\frac{dr^{2}}{1-kr^{2}/r_{0}^{2}}-r^{2}d\Omega^{2},\quad k=-1,0,1.\label{FLRW conf}
\end{equation}
Properties of the conformal metric $\bar{g}$ have been studied before,
e.g., by Deruelle and Sasaki \citep{Deruelle-Sasaki} and Dom\`enech
and Sasaki \citep{DomenechSasaki2016IJMPD}, implications of which
we discuss. Conformal frames have the light cone in common, so that
cosmic evolution and observation in the FLRW frame can be reinterpreted
in terms of gravitational time dilation in the static space of the
conformal frame $\bar{g}$, which is our main subject. 

One expects a cosmology of gravitational time dilation in the thermostatic
equilibrium of the conformal frame to deviate considerably from Big
Bang cosmology in the FLRW frame. Still, the two frames remain each
other's conformal dual, so the two cosmologies must be consistent.
The conformal frame has the distinct property of time-translation
symmetry, i.e., is maximally symmetric. This must have distinct implications
to cosmology in the expanding frame. The connection becomes transparent
when expressing the line element in conformal time $d\eta=a^{-1}dt$,
so that the metric assumes the static form, 

\begin{equation}
d\bar{s}^{2}=d\eta^{2}-\frac{dr^{2}}{1-kr^{2}/r_{0}^{2}}-r^{2}d\Omega^{2}.\label{static metric}
\end{equation}
Florides showed that the existence of a static representation constrains
solution space in the FLRW frame to spacetimes of constant curvature
\citep{Florides}. In particular, if density is positive, this regards
de Sitter spacetime ($k=-1,0,1)$. The constant energy density $\bar{\rho}_{\textrm{m}}=\rho_{\textrm{m}0}$
of ordinary matter in the conformal frame can then be considered to
act as cosmological constant, represented (in units where $\frac{8}{3}\pi G=c=1$)
by the constant spacetime curvature,
\begin{equation}
R_{0}^{-2}=\bar{\rho}_{\textrm{m}}=\rho_{\textrm{m}0}.\label{R0^-2=00003Drhom}
\end{equation}
From this assumption we derive properties of a nonempty de Sitter
cosmology in the curved time of the conformal frame. The Friedmann
equation is obtained as the sum of independent null solutions, i.e.,
a hyperbolic de Sitter universe arises from the equipartition of kinetic
energy of recessional and peculiar components of the gravitational
field on the light cone at the de Sitter horizon, where, for $ar=R_{0},$
one has, in both frames, the (conformally invariant) equilibrium condition
\begin{equation}
da^{2}/a^{2}=dr^{2}/r^{2}.\label{eq:da/a=00003Ddr/r}
\end{equation}

Due to time-translation symmetry, cosmic evolution in the conformal
frame seems to regard past and future only. This suggests that, regardless
of the model chosen, density parameters are fixed and must somehow
relate to the equilibrium state of the hyperbolic de Sitter universe,
therefore must be expressible in terms of the single independent parameter
$R_{0}^{-2}=\rho_{\textrm{m}0}$. In fact, a simple transformation
of the scale factor brings the hyperbolic de Sitter model into the
form of the basic $\Lambda\textrm{CDM}$ model. This transformation
gives exact expressions of concordance model parameters in terms of
the ordinary matter density $\rho_{\textrm{m}0}$. To introduce the
subject, we explore several relevant properties of the conformal frame
and of curved time in de Sitter universe. 

\subsection*{{\normalsize{}Properties of the conformal frame} }

In the maximal symmetry of the conformal frame in Eq.(\ref{FLRW conf})
one expects particle rest mass $\bar{m}$ to be constant. The conformal
factor $a^{-1}$ indeed restores the time-translation symmetry that
is lost in the expanding space of the FLRW frame. The authors of \citep{Deruelle-Sasaki,DomenechSasaki2016IJMPD}
show that mass $m$ in the FLRW frame transforms in the conformal
frame into 
\begin{equation}
\bar{m}=am.\label{eq:m_=00003Dam}
\end{equation}
This means that a presumed constant particle mass $m$ in expanding
space is in conflict with the notion of a constant particle mass $\bar{m}$
in the conformal frame. Turning the argument around: constant $\bar{m}$
in the time-translation symmetry of the conformal frame implies $m\propto a^{-1}$
in the expanding frame, so that $m$ evolves like cosmic temperature
$T\propto a^{-1}$. Then baryon density would evolve like radiation
density,
\begin{equation}
\rho_{\textrm{b}}=\rho_{\textrm{b0}}a^{-4}.\label{rhob=00003Da^-4}
\end{equation}
This suggests that the local energy density of the ordinary matter
content of baryons, photons and neutrinos evolves uniformly as 
\begin{equation}
\rho_{\textrm{m}}=\rho_{\textrm{b}}+\rho_{\gamma}+\rho_{\nu}=\rho_{\textrm{m0}}a(t)^{-4},\label{eq:rho_m=00003Da^-4}
\end{equation}
i.e., like the density of gravitational radiation associated with
the matter.

An evolving rest mass $\propto a^{-1}$ actually matches a version
of Mach's principle due to Schr\"odinger. In the Machian view, mass
is only due to interaction, therefore, is not an intrinsic property
of an object, as pointed out already by Berkeley, who questioned the
observability and physical relevance of a single object's mass in
otherwise empty space \citep{berkeley}. Schr\"odinger considered
the nonlocal mutual mass $\mu_{ij}\propto m_{i}m_{j}/\varphi_{0}R_{ij}$
between any two causally connected nominal masses $m_{i}$ and $m_{j}$
at separation $R_{ij}$, from which he retrieved Einstein's expression
of the anomalous perihelion precession \citep{Schroedinger,TelkampPhysRevD.94.043520,TelkampMassPhysRevD.98.063507}.
In the cosmological case one has on average $\left\langle R_{ij}\right\rangle \propto a$,
so that Schr\"odinger's nonlocal form of rest mass evolves identically
as $a^{-1}$.

We will not directly use these results, but independently retrieve
a radiation density, along with an equal vacuum energy density, from
analysis of the gravitational field of ordinary matter, subject to
constant curvature in the conformal frame. This involves several properties
related to curved time on the light cone at the horizon, which we
shall summarize first. 

\subsection*{{\normalsize{}Curved time at the de Sitter horizon}}

Recessional Doppler redshift is absent in the static space of the
conformal frame, so that gravitational redshift equals cosmological
redshift. This simplifies derivation of the distance-redshift relation
on the light cone in de Sitter universe, which can be obtained directly
from the constant radius $R_{0}=a(z)r(z)$ of the horizon. Given cosmological
redshift, i.e., $a(z)=(1+z)^{-1}$, integration of $dr(z)=-a(z)^{-2}R_{0}da(z)=R_{0}dz$
returns the linear distance-redshift relation 
\begin{equation}
r(z)=R_{0}z(r),\label{r=00003Dr0z}
\end{equation}
so that at the horizon redshift $z'\equiv z(r_{0})=1$ and the corresponding
scale factor $a'\equiv(1+z')^{-1}=\frac{1}{2}$. This expresses a
two times higher cosmic temperature $T'=T/a'$ at the horizon in the
conformal frame. Proper time $t'$ of a clock at the horizon appears
to run twice as fast as the proper time of the comoving observer,
i.e., $dt'=dt/a'=2dt$. As a result, the scale factor in the conformal
frame at the horizon of a flat de Sitter universe appears to evolve
exponentially faster (assuming at present $t'=t=0$), 
\begin{equation}
a(t')=e^{t'/R_{0}}=e^{2t/R_{0}}=a(t)^{2}.\label{a(t')=00003Da(t)^2}
\end{equation}
This exponential relation {[}in general $a\bigl((1+z)t\bigr)=a(t)^{1+z}${]}
is actually required for invariance of spacetime curvature in the
curved time of the conformal frame, since
\begin{equation}
\frac{\mathring{a}}{a}\equiv\frac{da(t')}{a(t')dt'}=\frac{da(t)^{2}}{a(t)^{2}2dt}=\frac{da(t)}{a(t)dt}\equiv\frac{\dot{a}}{a},\label{Invar curv}
\end{equation}
where the ring symbol $\mathring{}$ denotes differentiation with
respect to proper time $t'$ in the conformal frame at the horizon.
Using Eq.(\ref{eq:da/a=00003Ddr/r}), this gives for null geodesics
the identities
\begin{equation}
\frac{\dot{r}^{2}}{r^{2}}=\frac{\dot{a}^{2}}{a^{2}}=\frac{\mathring{a}^{2}}{a^{2}}=\frac{\mathring{r}^{2}}{r^{2}}.\label{da/a^2=00003Ddr/r^2}
\end{equation}

Ingoing ($-$) and outgoing ($+$) null geodesics satisfy
\begin{equation}
\frac{\dot{R}_{\pm}^{2}}{R^{2}}=\frac{\dot{a}^{2}}{a^{2}}+2\frac{\dot{a}}{a}\frac{\dot{r}_{\pm}}{r}+\frac{\dot{r}_{\pm}^{2}}{r^{2}},\label{dR/R=00003Dda/a+dr/r}
\end{equation}
where the alternating cross term vanishes in the ensemble average.
With Eq.(\ref{da/a^2=00003Ddr/r^2}), and simplifying $dr^{2}\equiv dr_{\pm}^{2}$,
this gives for the ensemble average
\begin{equation}
\frac{\dot{R}^{2}}{R^{2}}=\frac{\dot{a}^{2}}{a^{2}}+\frac{\dot{r}^{2}}{r^{2}}=2\frac{\dot{a}^{2}}{a^{2}}=2\frac{\mathring{a}^{2}}{a^{2}}=\frac{\mathring{a}^{2}}{a^{2}}+\frac{\mathring{r}^{2}}{r^{2}}=\frac{\mathring{R}^{2}}{R^{2}}.\label{dR/Rdt'=00003DdR/Rdt=00003Dda/adt}
\end{equation}
It shows that one can distinguish between the Hubble parameter in
the original definition, expressing recessional speed $H\equiv\dot{a}/a$,
and the Hubble parameter $\tilde{H}\equiv\dot{R}/R$ expressing (the
square root of) total energy density. The difference in magnitude
$\dot{R}/R=\sqrt{2}\dot{a}/a$ may be part of what is known as the
Hubble tension, as indicated further on. 

\subsection*{{\normalsize{}Nonlocal energy density of a nonempty de Sitter universe}}

Given invariance of spacetime curvature, the ingoing and outgoing
null geodesics of the gravitational field at two antipodal points
at the de Sitter horizon in the conformal frame satisfy, in terms
of the local time $t'$, the 4 distinct equations
\begin{equation}
\mathring{R}_{\pm\pm}=\pm\left(\frac{R(t')}{R_{0}}\pm1\right).\label{eq:Null de Sitter}
\end{equation}
For symmetry of the field at the 2 antipodal points in each of $n$=3
spatial dimensions, the density of, respectively, ingoing ($-$) and
outgoing ($+$) fields amounts to 
\begin{equation}
\frac{\mathring{R}_{\pm}^{2}}{R^{2}}=\frac{6}{R_{0}^{2}}\left(1\pm\frac{R_{0}}{R(t')}\right)^{2}.\label{Fr +/- single t'}
\end{equation}
With cross terms canceling in the ensemble of null rays, one obtains,
in terms of dilated time $t'$, the Friedmann equation of a nonempty
hyperbolic de Sitter universe, 
\begin{equation}
\frac{\mathring{R}^{2}}{R^{2}}=\frac{12}{R_{0}^{2}}\left(1+\frac{R_{0}^{2}}{R(t')^{2}}\right)=\frac{12}{R_{0}^{2}}\left(1+a(t')^{-2}\right).\label{Fr R total t'}
\end{equation}
Like before, the densities in Eq.(\ref{Fr R total t'}) transform
invariantly to the frame of the observer by the substitutions $\mathring{R}^{2}/R^{2}=\dot{R}^{2}/R^{2}$
and $a(t')=a(t)^{2}$, so that the total density in terms of comoving
time $t$ is expressed by the Friedmann equation 
\begin{equation}
\tilde{H}(t)^{2}\equiv\frac{\dot{R}^{2}}{R^{2}}=\frac{12}{R_{0}^{2}}\left(1+a(t)^{-4}\right)=\frac{12}{R_{0}^{2}}\left(1+\frac{R_{0}^{4}}{R(t)^{4}}\right).\label{Fr R total t}
\end{equation}
This is where curvature energy density $12R_{0}^{-2}a(t')^{-2}$ appears
in the frame of the observer as gravitational radiation energy density
$12R_{0}^{-2}a(t)^{-4}$. Notice that the equally large constant 'vacuum'
energy density $4nR_{0}^{-2}=12R_{0}^{-2}$ in $n$=3 dimensional
space matches the scalar curvature $d(d-1)R_{0}^{-2}$ of the $d=n+1$
dimensional de Sitter spacetime (interestingly, in $n=3$ dimensional
space only). 

The nonempty de Sitter universe has a total nonlocal energy density
$24\rho_{\textrm{m0}}$, corresponding to an exact (relative) local
energy density of ordinary matter
\begin{equation}
\Omega_{\textrm{m}}=\frac{1}{24}=0.04166...\;.\label{Ohmm}
\end{equation}
This can be validated using the baryon density estimate $(\Omega_{\textrm{b}}h^{2})_{\textrm{Planck}}=0.0224\pm0.0001$
(Planck 2018 \citep{Planck2018A&A}). Without introducing dark components,
it predicts a total density $24\Omega_{\textrm{b}}h^{2}$, corresponding
to a Hubble constant 
\begin{equation}
\tilde{h}_{\textrm{Planck}}=\sqrt{24(\Omega_{\textrm{b}}h^{2})_{\textrm{Planck}}}=0.7332\pm0.0016,\label{h_24b}
\end{equation}
in agreement with the presently most accurate distance ladder estimates
of the Hubble constant, e.g., $0.7304\pm0.0104$ by the SH0ES collaboration
\citep{SHOES2022ApJ...934L...7R}, $0.732\pm0.013$ by Riess \textit{et
al.} \citep{Riess2021ApJ...908L...6R}, or $0.733\pm0.018$ by Wong
\textit{et al.} \citep{Holicow}. The BBN estimate of baryon density
$(\Omega_{\textrm{b}}h^{2})_{\textrm{BBN}}=0.02166\pm0.00026$ by
Cooke \textit{et al.} \citep{BBNCooke_2018} predicts 
\begin{equation}
\tilde{h}_{\textrm{BBN}}=0.7210\pm0.0043,\label{h~BBN}
\end{equation}
still within range of the cited $H_{0}$ estimates. The agreement
suggests that distance ladder estimates represent total energy density
$\tilde{H}_{0}^{2}=24\rho_{\textrm{m0}}$ of the hyperbolic de Sitter
universe. 

\subsection*{{\normalsize{}Predicting concordance model parameters}}

The blackbody spectrum of the CMB points at thermal equilibrium, stationarity
and equipartition of kinetic energy. This seems to naturally fit the
hyperbolic de Sitter solution in Eq.(\ref{da/adt}). In fact, the
basic $\Lambda\textrm{CDM}$ model with energy densities of vacuum
and pressureless matter has a solution of the form $\hat{A}^{\nicefrac{1}{3}}\textrm{sinh}(t/\hat{R}_{0})^{\nicefrac{2}{3}}$,
so is within the family of hyperbolic de Sitter solutions. Viability
of the hyperbolic de Sitter model can therefore be investigated by
transformation into $\Lambda\textrm{CDM}$ form, as a way to predict
concordance model parameters, as follows. 

The anisotropy of CMB temperature provides a measure of matter density
fluctuations. These fluctuations can be decomposed into spherical
harmonic acoustic oscillations of the matter fluid, which represent
peculiar kinetic energy density. In thermodynamic equilibrium one
expects, in accordance with Eq.(\ref{da/a^2=00003Ddr/r^2}), the peculiar
kinetic energy density to match recessional kinetic energy density,
which presents the Hubble parameter (in agreement with the original
definition)
\begin{equation}
H(t)^{2}\equiv\frac{\dot{a}^{2}}{a^{2}}=\frac{6}{R_{0}^{2}}\left(1+a(t)^{-4}\right)=\frac{\dot{r}^{2}}{r^{2}}.\label{da/adt}
\end{equation}
This Friedmann equation has solution 
\begin{equation}
a(t)=\textrm{sinh}(2\sqrt{6}t/R_{0})^{\nicefrac{1}{2}},\label{a(t)}
\end{equation}
which, like the flat de Sitter solution in Eq.(\ref{a(t')=00003Da(t)^2}),
satisfies the condition $a(t)^{2}=a(t')=\textrm{sinh}(\sqrt{6}t'/R_{0})$
for invariance of densities $\dot{a}^{2}/a^{2}=\mathring{a}^{2}/a^{2}$.
A curious implication of time-translation symmetry in the conformal
frame is that even the age parameter $t_{0}$ is a constant. For $a(t_{0})=1$
one obtains the constant age 
\begin{equation}
t_{0}=R_{0}\textrm{asinh}(1)/2\sqrt{6},\label{eq:t0}
\end{equation}
evidently not representing elapsed time, but constant curvature. This
seemingly problematic notion of age still is sensible in the conformal
frame, where, relative to $dt$, the propagation of proper (conformal)
time $d\eta=dt/a(t)$ slows down, so that age becomes a constant $t_{0}$
in terms of the continuously decreasing present clock rate. In this
way the finite age Big Bang scenario in the expanding frame still
fits an infinitely old steady state universe in terms of proper time
$d\eta$ in the conformal frame. 

$H^{2}=\dot{a}^{2}/a^{2}$ represents only half the total energy density
$\tilde{H}^{2}=\dot{R}^{2}/R^{2}=2\dot{a}^{2}/a^{2}$, at a correspondingly
small value of the Hubble constant, $h=\tilde{h}/\sqrt{2}\approx0.5184$,
i.e., significantly below concordance model estimates $\hat{h}\approx0.674$.
However, the Hubble constant is a model dependent parameter, as noted
in \citep{Planck2018A&A} and like shown in the following transformation
into $\Lambda\textrm{CDM}$ form. 

The $\Lambda\textrm{CDM}$ form of Friedmann equation (\ref{da/adt})
can be obtained by (a noninvariant) transformation of the scale factor
$a(t)\rightarrow\hat{a}(t)$, where
\begin{equation}
\hat{a}(t)^{3}/a(t)^{4}=\textrm{const}.\label{eq:a^3=00003DCa^4}
\end{equation}
It brings radiation density $\propto a^{-4}$ into the form of pressureless
matter density $\propto\hat{a}^{-3}$. But it also distorts parameters,
like the Hubble parameter, since 
\begin{equation}
\hat{H}\equiv\frac{\dot{\hat{a}}}{\hat{a}}=\frac{4}{3}\frac{\dot{a}}{a}=\frac{4}{3}H,\label{da^/a^=00003D4/3da/a}
\end{equation}
causing the Hubble constant to appear a factor $4/3$ larger in the
concordance $\Lambda\textrm{CDM}$ model, i.e.,
\begin{equation}
\hat{h}=\frac{4}{3}h=\frac{4}{3}\sqrt{12}R_{0}^{-1}.\label{eq:h^}
\end{equation}
With the Planck 2018 estimate of the baryon density $(\Omega_{\textrm{b}}h^{2})_{\textrm{Planck}}=0.0224\pm0.0001$
\citep{Planck2018A&A}, this predicts $\hat{h}_{\textrm{b,Planck}}=\frac{4}{3}\sqrt{12(\Omega_{\textrm{b}}h^{2})_{\textrm{Planck}}}=0.6913\pm0.002$.
Alternatively, the BBN estimate $(\Omega_{\textrm{b}}h^{2})_{\textrm{BBN}}=0.02166\pm0.00026$
\citep{BBNCooke_2018} gives 
\begin{equation}
\hat{h}_{\textrm{BBN}}=\frac{4}{3}\sqrt{12(\Omega_{\textrm{b}}h^{2})_{\textrm{BBN}}}=0.6798\pm0.0041,\label{h^BBN}
\end{equation}
which matches the Planck 2018 estimate $\hat{h}_{\textrm{Planck}}=0.674\pm0.005$
within confidence limits. Also notice that the exact ratio
\begin{equation}
\hat{h}/\tilde{h}=\frac{4}{3}/\sqrt{2}=\sqrt{8/9}=0.9428...\;,\label{Hub tens}
\end{equation}
is within range of what is known as the Hubble tension, roughly $(67.4\pm0.5)/(73\pm2)\sim0.925\pm0.030$. 

Given Eqs.(\ref{eq:a^3=00003DCa^4}) and (\ref{da^/a^=00003D4/3da/a}),
the Friedmann equation of the basic $\Lambda\textrm{CDM}$ model can
be written

\begin{equation}
\frac{\dot{\hat{a}}^{2}}{\hat{a}^{2}}=\frac{16}{9}\frac{12}{R_{0}^{2}}\left(\hat{\Omega}_{\Lambda}+\hat{\Omega}_{\textrm{m}}\hat{a}^{-3}\right),\label{LCDM}
\end{equation}
subject to $\hat{\Omega}_{\Lambda}+\hat{\Omega}_{\textrm{m}}=1$,
where the density ratio $\hat{A}\equiv\hat{\Omega}_{\textrm{m}}/\hat{\Omega}_{\Lambda}$
is to be determined. The effect of the (noninvariant) transformation
$a(t)\rightarrow\hat{a}(t)$ on $\hat{A}$ is indirect, since the
density ratio is also connected with the constant age parameter $t_{0}$
of Eq.(\ref{eq:t0}). It is therefore convenient to decompose the
transformation into two steps, $a(t)\rightarrow\bar{a}(t)\rightarrow\hat{a}(t)$,
where both steps have a known effect on the density ratio $\hat{A}$.
Given $a(t)=A^{\nicefrac{1}{4}}\textrm{sinh}(2\sqrt{6}t/R_{0})^{\nicefrac{1}{2}}$
and $t_{0}=R_{0}\textrm{asinh}(1)/2\sqrt{6}$, we (provisionally)
scale the time coordinate by a factor $\frac{4}{3}$, i.e.,
\begin{equation}
\bar{a}(t)=\bar{A}^{\nicefrac{1}{4}}\textrm{sinh}(\frac{4}{3}2\sqrt{6}t/R_{0})^{\nicefrac{1}{2}}.\label{a-(t)}
\end{equation}
Like with scaling of the exponent, this noninvariant transformation
raises the Hubble constant to $\bar{H}_{0}=\frac{4}{3}H_{0}$, while,
for $\bar{a}(t_{0})=1$, one obtains the exact density ratio 
\begin{equation}
\bar{A}=\textrm{sinh}(\frac{4}{3}\textrm{asinh}(1))^{-2}=0.4660...\;.\label{A^}
\end{equation}
The second step is \textsl{invariant} transformation $\bar{a}(t)\rightarrow\hat{a}(t)$
of the intermediate solution into the $\Lambda\textrm{CDM}$ solution,
which preserves densities, i.e., both total density $\hat{H}_{0}^{2}=\bar{H}_{0}^{2}=\frac{16}{9}H_{0}^{2}$
and density ratio $\hat{A}=\bar{A}=0.4660...\;$. This involves scaling
of the exponents in $\bar{a}(t)$ by the factor $\frac{4}{3}$, along
with opposite scaling of the time coordinate by the inverse factor
$\frac{3}{4}$ (which undoes the initial scaling of the time coordinate).
Hence,
\begin{equation}
\hat{a}(t)=\hat{A}^{\nicefrac{1}{3}}\textrm{sinh}(2\sqrt{6}t/R_{0})^{\nicefrac{2}{3}}.\label{a^(t)}
\end{equation}
Projection of the equilibrium state of the model in Eq.(\ref{da/adt})
onto the present state of the $\Lambda\textrm{CDM}$ model thus gives
an exact (non-coincidental) value of the relative energy densities
of dust and vacuum in $\Lambda\textrm{CDM}$, i.e.,
\begin{equation}
\hat{\Omega}_{\textrm{m}}=1-\hat{\Omega}_{\Lambda}=\frac{\hat{A}}{1+\hat{A}}=\frac{1}{\textrm{sinh}(\frac{4}{3}\textrm{asinh}(1))^{2}+1}=0.3179...\label{O^hmm}
\end{equation}
within confidence limits of the Planck 2018 estimate $\Omega_{\textrm{m}}^{\textrm{Planck}}=0.315\pm0.007$
\citep{Planck2018A&A}. 

It shows that, without introducing dark components, accurate predictions
of concordance model parameters can be derived from the gravitational
field of the known ordinary matter content only. The close agreement
of $\hat{\Omega}_{\textrm{m}}=1-\hat{\Omega}_{\Lambda}$ with observation
seems indirect evidence of the underlying equilibrium of recessional
and peculiar density components of the gravitational field associated
with ordinary matter in the maximal symmetry of the conformal frame.

\bibliographystyle{apsrev}

\begin{thebibliography}{15}
\expandafter\ifx\csname natexlab\endcsname\relax\def\natexlab#1{#1}\fi
\expandafter\ifx\csname bibnamefont\endcsname\relax
  \def\bibnamefont#1{#1}\fi
\expandafter\ifx\csname bibfnamefont\endcsname\relax
  \def\bibfnamefont#1{#1}\fi
\expandafter\ifx\csname citenamefont\endcsname\relax
  \def\citenamefont#1{#1}\fi
\expandafter\ifx\csname url\endcsname\relax
  \def\url#1{\texttt{#1}}\fi
\expandafter\ifx\csname urlprefix\endcsname\relax\def\urlprefix{URL }\fi
\providecommand{\bibinfo}[2]{#2}
\providecommand{\eprint}[2][]{\url{#2}}

\bibitem[{\citenamefont{Poincare}(1914)}]{PoincareScienceMethod}
\bibinfo{author}{\bibfnamefont{H.}~\bibnamefont{Poincare}},
  \emph{\bibinfo{title}{Science and method; The Relativity of Space}}
  (\bibinfo{publisher}{T. Nelson London}, \bibinfo{year}{1914}).

\bibitem[{\citenamefont{Hoyle}(1948)}]{Hoyle}
\bibinfo{author}{\bibfnamefont{F.}~\bibnamefont{Hoyle}}, \bibinfo{journal}{Mon.
  Not. Roy. Astron. Soc.} \textbf{\bibinfo{volume}{108}}, \bibinfo{pages}{372}
  (\bibinfo{year}{1948}).

\bibitem[{\citenamefont{Bondi and Gold}(1948)}]{BondiGold}
\bibinfo{author}{\bibfnamefont{H.}~\bibnamefont{Bondi}} \bibnamefont{and}
  \bibinfo{author}{\bibfnamefont{T.}~\bibnamefont{Gold}},
  \bibinfo{journal}{Mon. Not. Roy. Astron. Soc.}
  \textbf{\bibinfo{volume}{108}}, \bibinfo{pages}{252} (\bibinfo{year}{1948}).

\bibitem[{\citenamefont{Deruelle and Sasaki}(2011)}]{Deruelle-Sasaki}
\bibinfo{author}{\bibfnamefont{N.}~\bibnamefont{Deruelle}} \bibnamefont{and}
  \bibinfo{author}{\bibfnamefont{M.}~\bibnamefont{Sasaki}}, in
  \emph{\bibinfo{booktitle}{Cosmology, Quantum Vacuum and Zeta Functions}}
  (\bibinfo{publisher}{Springer Berlin Heidelberg}, \bibinfo{address}{Berlin,
  Heidelberg}, \bibinfo{year}{2011}), pp. \bibinfo{pages}{247--260}, ISBN
  \bibinfo{isbn}{978-3-642-19760-4}.

\bibitem[{\citenamefont{{Dom{\`e}nech} and
  {Sasaki}}(2016)}]{DomenechSasaki2016IJMPD}
\bibinfo{author}{\bibfnamefont{G.}~\bibnamefont{{Dom{\`e}nech}}}
  \bibnamefont{and} \bibinfo{author}{\bibfnamefont{M.}~\bibnamefont{{Sasaki}}},
  \bibinfo{journal}{International Journal of Modern Physics D}
  \textbf{\bibinfo{volume}{25}}, \bibinfo{eid}{1645006} (\bibinfo{year}{2016}).

\bibitem[{\citenamefont{{Florides}}(1980)}]{Florides}
\bibinfo{author}{\bibfnamefont{P.~S.} \bibnamefont{{Florides}}},
  \bibinfo{journal}{General Relativity and Gravitation}
  \textbf{\bibinfo{volume}{12}}, \bibinfo{pages}{563} (\bibinfo{year}{1980}).

\bibitem[{\citenamefont{Berkeley}(1992)}]{berkeley}
\bibinfo{author}{\bibfnamefont{G.}~\bibnamefont{Berkeley}},
  \emph{\bibinfo{title}{{De Motu and The Analyst}}}
  (\bibinfo{publisher}{Springer. New York}, \bibinfo{year}{1992}).

\bibitem[{\citenamefont{Schr{\"o}dinger}(1925)}]{Schroedinger}
\bibinfo{author}{\bibfnamefont{E.}~\bibnamefont{Schr{\"o}dinger}},
  \bibinfo{journal}{Annalen der Physik} \textbf{\bibinfo{volume}{382}},
  \bibinfo{pages}{325} (\bibinfo{year}{1925}), \bibinfo{note}{{English
  translation J.B. Barbour in \textit {Mach's Principle: From Newton's Bucket
  to Quantum Gravity} (Birkenhauser, Boston 1995)}}.

\bibitem[{\citenamefont{{Telkamp}}(2016)}]{TelkampPhysRevD.94.043520}
\bibinfo{author}{\bibfnamefont{H.}~\bibnamefont{{Telkamp}}},
  \bibinfo{journal}{Phys. Rev. D} \textbf{\bibinfo{volume}{94}},
  \bibinfo{eid}{043520} (\bibinfo{year}{2016}).

\bibitem[{\citenamefont{{Telkamp}}(2018)}]{TelkampMassPhysRevD.98.063507}
\bibinfo{author}{\bibfnamefont{H.}~\bibnamefont{{Telkamp}}},
  \bibinfo{journal}{Phys. Rev. D} \textbf{\bibinfo{volume}{98}},
  \bibinfo{eid}{063507} (\bibinfo{year}{2018}).

\bibitem[{\citenamefont{Aghanim et~al.}(2020)}]{Planck2018A&A}
\bibinfo{author}{\bibfnamefont{N.}~\bibnamefont{Aghanim}} \bibnamefont{et~al.}
  (\bibinfo{collaboration}{Planck collaboration}), \bibinfo{journal}{Astron.
  Astrophys.} \textbf{\bibinfo{volume}{641}}, \bibinfo{pages}{A6}
  (\bibinfo{year}{2020}).

\bibitem[{\citenamefont{{Riess} et~al.}(2022)\citenamefont{{Riess}, {Yuan},
  {Macri}, {Scolnic}, {Brout}, {Casertano}, {Jones}, {Murakami}, {Anand},
  {Breuval} et~al.}}]{SHOES2022ApJ...934L...7R}
\bibinfo{author}{\bibfnamefont{A.~G.} \bibnamefont{{Riess}}},
  \bibinfo{author}{\bibfnamefont{W.}~\bibnamefont{{Yuan}}},
  \bibinfo{author}{\bibfnamefont{L.~M.} \bibnamefont{{Macri}}},
  \bibinfo{author}{\bibfnamefont{D.}~\bibnamefont{{Scolnic}}},
  \bibinfo{author}{\bibfnamefont{D.}~\bibnamefont{{Brout}}},
  \bibinfo{author}{\bibfnamefont{S.}~\bibnamefont{{Casertano}}},
  \bibinfo{author}{\bibfnamefont{D.~O.} \bibnamefont{{Jones}}},
  \bibinfo{author}{\bibfnamefont{Y.}~\bibnamefont{{Murakami}}},
  \bibinfo{author}{\bibfnamefont{G.~S.} \bibnamefont{{Anand}}},
  \bibinfo{author}{\bibfnamefont{L.}~\bibnamefont{{Breuval}}},
  \bibnamefont{et~al.}, \bibinfo{journal}{Astrophys. J.}
  \textbf{\bibinfo{volume}{934}}, \bibinfo{eid}{L7} (\bibinfo{year}{2022}),
  \eprint{2112.04510}.

\bibitem[{\citenamefont{{Riess} et~al.}(2021)}]{Riess2021ApJ...908L...6R}
\bibinfo{author}{\bibfnamefont{A.~G.} \bibnamefont{{Riess}}}
  \bibnamefont{et~al.}, \bibinfo{journal}{Astrophys. J. Letters}
  \textbf{\bibinfo{volume}{908}}, \bibinfo{eid}{L6} (\bibinfo{year}{2021}).

\bibitem[{\citenamefont{Wong et~al.}(2019)}]{Holicow}
\bibinfo{author}{\bibfnamefont{K.~C.} \bibnamefont{Wong}} \bibnamefont{et~al.},
  \bibinfo{journal}{Mon. Not. Roy. Astron. Soc.}
  \textbf{\bibinfo{volume}{498}}, \bibinfo{pages}{1420} (\bibinfo{year}{2019}).

\bibitem[{\citenamefont{Cooke et~al.}(2018)\citenamefont{Cooke, Pettini, and
  Steidel}}]{BBNCooke_2018}
\bibinfo{author}{\bibfnamefont{R.~J.} \bibnamefont{Cooke}},
  \bibinfo{author}{\bibfnamefont{M.}~\bibnamefont{Pettini}}, \bibnamefont{and}
  \bibinfo{author}{\bibfnamefont{C.~C.} \bibnamefont{Steidel}},
  \bibinfo{journal}{The Astrophysical Journal} \textbf{\bibinfo{volume}{855}},
  \bibinfo{pages}{102} (\bibinfo{year}{2018}).

\end{thebibliography}

\end{document}